\documentclass[a4paper,fleqn]{cas-sc}

\graphicspath{{figures/}} 
\usepackage[authoryear]{natbib}
\usepackage{algorithm2e}
\usepackage{subfloat}
\usepackage{amssymb}
\usepackage{amsmath}
\usepackage{subcaption}
\usepackage{placeins}
\usepackage{textcomp}
\usepackage{xcolor}

\newcommand{\Tb}{\boldsymbol{T}}

\newcommand{\lambdab    }{\boldsymbol{\lambda}}
\newcommand{\thetab    }{\boldsymbol{\theta}}
\newcommand{\komegasst}{$k$-$\omega$ SST}
\newcommand{\frozen}{$k$-corrective-frozen-RANS}

\def\tsc#1{\csdef{#1}{\textsc{\lowercase{#1}}\xspace}}
\tsc{WGM}
\tsc{QE}
\tsc{EP}
\tsc{PMS}
\tsc{BEC}
\tsc{DE}

\begin{document}
\let\WriteBookmarks\relax
\def\floatpagepagefraction{1}
\def\textpagefraction{.001}
\shorttitle{Customized RANS closures for bi-fidelity optimization}
\shortauthors{Y Zhang et~al.}

\title[mode = title]{Customized data-driven RANS closures for bi-fidelity LES--RANS optimization}
\tnotemark[1]

\tnotetext[1]{This document is the results of the research project funded by Innovation Foundation for Doctor Dissertation of Northwestern Polytechnical University under grant No. CX201801 and China Scholarship Council No. 201706290083.}


\author[1]{Yu Zhang}
\ead{zhangyu91@mail.nwpu.edu.cn}
\author[2]{Richard P. Dwight}
\cormark[1]
\ead{r.p.dwight@tudelft.nl}
\author[2]{Martin Schmelzer}
\ead{m.schmelzer@tudelft.nl}
\author[2]{Javier F. G\'{o}mez}
\ead{jfatoug@gmail.com}
\author[2]{Stefan Hickel}
\ead{s.hickel@tudelft.nl}
\author[1]{Zhong-hua Han}
\ead{hanzh@nwpu.edu.cn}

\address[1]{Institute of Aerodynamic and Multidisciplinary Design Optimization, National Key Laboratory of Science and Technology on Aerodynamic Design and Research, School of Aeronautics, Northwestern Polytechnic University, Youyi West Road 127, 710072 Xi'an, People's Republic of China}
\address[2]{Aerodynamics Group, Faculty of Aerospace Engineering, Delft University of Technology, Kluyverweg 2, 2629 HS Delft, The Netherlands}

\cortext[cor1]{Corresponding author}

\begin{abstract}
  Multi-fidelity optimization methods promise a high-fidelity optimum at a cost only slightly greater than a low-fidelity optimization.  This promise is seldom achieved in practice, due to the requirement that low- and high-fidelity models correlate well. In this article, we propose an efficient bi-fidelity shape optimization method for turbulent fluid-flow applications with Large-Eddy Simulation (LES) and Reynolds-averaged Navier-Stokes (RANS) as the high- and low-fidelity models within a hierarchical-Kriging surrogate modelling framework.  Since the LES--RANS correlation is often poor, we use the full LES flow-field at a single point in the design space to derive a {\it custom-tailored RANS closure model} that reproduces the LES at that point.  This is achieved with machine-learning techniques, specifically sparse regression to obtain high corrections of the turbulence anisotropy tensor and the production of turbulence kinetic energy as functions of the RANS mean-flow.  The LES--RANS correlation is dramatically improved throughout the design-space.  We demonstrate the effectiveness and efficiency of our method in a proof-of-concept shape optimization of the well-known periodic-hill case.  Standard RANS models perform poorly in this case, whereas our method converges to the LES-optimum with only two LES samples.
\end{abstract}


\maketitle

\section{Introduction}

Numerical fluid-dynamic shape-optimization is an increasingly central tool in engineering practice.  Typically Reynolds-Averaged Navier-Stokes (RANS) is used as the fluid model, due to its acceptable computational cost and accuracy.  However in many important situations, the physics demands scale-resolving simulations of turbulence such as large-eddy simulation (LES).  This includes, for instance, designs where separation occurs; junctions flows with complex turbulence behaviour~\citep{Simpson2001,Belligoli_2019}; and novel boundary layer flow control applications.  LES is often applied for a final analysis and validation, but its high computational cost precludes its use within the design loop.

Multi-fidelity optimization (MFO) methods are supposed to address exactly this dilemma. In MFO, low-fidelity but cheap-to-evaluate physics models are used to explore the design-space rapidly; their predictions are then corrected by a few expensive high-fidelity simulations. One such class of methods are the multi-fidelity surrogate modelling methods~\citep{Han2012b}, combined with a suitable sampling criteria~\citep{Jones1998}. 

There has been significant work on multi-fidelity surrogate modelling methods, almost all based on Gaussian-process models for the surrogate (i.e.\ Kriging), differing mainly in the manner in which the high-fidelity correct the lower.  \citet{Haftka1991} and \citet{Chang1993} developed a variable-fidelity Kriging model using a multiplicative ``bridge function'' to correct the low-fidelity model. \citet{Gano2005} developed a hybrid bridge function method, which uses a second Kriging model to scale the low-fidelity model. \citet{Han2013} combined gradient information and a generalized hybrid bridge function.  Cokriging was originally proposed in the geostatistics community by \citet{Journel1978} and then extended to deterministic computer experiments by \citet{Kennedy2000}. \citet{Han2012a} proposed an improved version of cokriging, which can be built in one step, and a hierarchical Kriging (HK) model \citep{Han2012b}, appears to be as simple and robust as the correction-based method and as accurate as cokriging method.

However, the success of all these approaches relies heavily on the correlation between the high- and low-fidelity models~\citep{Han2020,Zhang2018}. Since models with higher fidelity typically have also significantly higher numerical costs, in practice the high-fidelity model is chosen as the cheapest model that predicts the relevant physics. Necessarily, the low-fidelity model will lack some of the relevant physics, leading to a poor correlation between the models, which limits the practical application of MFO methods.  To this end, \citet{Gomez2020} proposed a novel multi-fidelity optimization method, injecting LES Reynolds stress statistics into RANS model to improve them, and then using both in a bi-fidelity optimization.

In separate developments of the past few years, data-driven methods for turbulence modelling based on supervised machine learning have been introduced to improve RANS predictions when reference data is available \citep{Xiao2019,Duraisamy2019,Durbin2018,Kaandorp_2020}.  Notably, \citet{Parish2016} introduced a local multiplicative term to correct the turbulence production in the $k$-equation of the $k-\omega$ model.  This term was chosen by solving an inverse problem to match LES reference data, and was then used to train a Gaussian process to correct the baseline turbulence model.  \citet{Ling2016} trained a deep neural network to predict turbulence anisotropy $a_{ij}$ and replace the baseline turbulence model.  An alternative is to use random-forests for the same task~\citep{Wang2016,Kaandorp_2020}. Even though these approaches generalize the linear eddy-viscosity concept, they generate complex black-box closure models.  More promising are approaches that generate compact explicit expressions for models.  Notable are gene-expression programming (GEP) \citep{Weatheritt2016,Weatheritt2017}, and deterministic sparse regression of turbulence anisotropy (SpaRTA) \citep{Schmelzer2020}.  These methods generate models that can be rapidly implemented in existing CFD codes and evaluated at every iteration of a RANS solution, and potentially inspected to identify the physical mechanisms influencing the flow.  The SpaRTA approach has shown the ability to consistently reproduce specified LES mean-flows in a RANS solver, using only three to five additional non-linear closure terms.  It is important to understand that -- at their present level of development -- these methods do not generate general-purpose turbulence models, but rather deliver corrections for flows similar to the flow they were trained on. Thus, for flows with a high degree of similarity, e.g. modifications in the geometry, the methods generate models effectively customized to the flow at hand.

The key original contribution of this paper is a highly-efficient bi-fidelity optimization procedure, using LES as the high-fidelity model, and {\it data-enhanced} RANS as the low-fidelity model.  In outline, we:
\begin{enumerate}
\item Perform a single LES simulation at the baseline geometry $\psi_0$, $i=0$;
\item Use the full-field LES data to generate a customized RANS model matching the LES at $\psi_i$;
\item Sample this custom RANS model throughout the design-space (cheap);
\item Combine custom-RANS and LES results in a bi-fidelity surrogate;
\item Perform a new LES at the point of maximum expected improvement $\psi_{i+1}$ (expensive);
\item Generate new custom RANS model(s), based on all LES data available so far;
\item $i \leftarrow i+1$, goto 3.
\end{enumerate}
The key observation is that the custom RANS model correlates well with LES in a region of the design-space for which the LES training data is informative.  Provided the physical processes occurring do not significantly change, the data-driven RANS remains a viable model.  Also, whereas typical multi-fidelity methods use only the objective function from the high-fidelity simulation, our approach uses much more information.  There are many variations on the basic pattern described above, for example: Which LES samples are used for training in Step 6?; What criteria are used for sampling RANS in Step 3, and LES in Step 5?.

In the following we demonstrate the method on a proof-of-concept optimization problem: a generalized periodic-hill (P-H) geometry, in which the steepness of the hills is varied~\citep{Gomez2018,Gomez2020}. The baseline periodic-hill flow (as proposed by \citet{Mellen2000} based on an experimental study by \citet{Almeida1993}) is notoriously for poor performance of essentially all standard RANS models, due to its sensitivity on the flow separation location and large-scale low-frequency dynamics. We use a hierarchical Kriging bi-fidelity surrogate model, with an EGO-like sampling procedure for the LES~\citep{Jones1998,Zhang2018}.  The custom RANS model is sampled on a uniform grid, and updated at each step with the latest LES data.  To achieve the LES optimal solution we require only two LES samples (and one further sample to verify the result).  Our method therefore offers a promising path towards LES-quality optimization with only a handful of LES samples.
 
The paper is structured as follows: in Section 2 we describe the computational setting of the baseline RANS and LES simulations.  Section 3 briefly describes SpaRTA, the machine-learning method used in this work to generate data-driven RANS models.  SpaRTA is demonstrated on the baseline periodic-hill case \citep{Mellen2000}.  Our proposed bi-fidelity optimization procedure is described in detail, and applied to the generalized P-H optimization case with the hill width as a design variable in Section 4.

\section{Baseline incompressible LES and RANS simulations}

The solvers used here for LES and RANS are both finite-volume methods, but otherwise numerically distinct - we describe both briefly.  For the standard periodic-hill test-case we compare our LES results with the experimental data of \citet{Rapp2009} and well-resolved LES reference data from \citet{Breuer2009}, and demonstrate excellent agreement.  We show that RANS results obtained with a standard \komegasst{} model are poor, which is in agreement with literature.

\subsection{Baseline Reynolds-Averaged Navier-Stokes with \komegasst{}}
\label{s:rans}

Assuming incompressible flow with constant fluid density equal to unity, the steady RANS equations are
\begin{align} \label{RANS}
\frac{\partial U_i}{\partial x_i} &= 0, \\ \nonumber
U_j \frac{\partial U_i}{\partial x_j} &= \frac{\partial}{\partial x_j} \left[ -\delta_{ij} {P} + \nu \frac{\partial U_i}{\partial x_j}- \tau_{ij} \right] + \delta_{i1} f,
\end{align}
where $U_i$ with $i,j \in \lbrace 1,2,3 \rbrace$ are the components of the mean-flow velocity, $P$ is the mean pressure, and $\nu$ is the kinematic viscosity.  The volume forcing $f$ serves to drive the flow through the doubly periodic domain.  The effect of turbulence on the momentum equation is confined to the Reynolds stress tensor $\tau_{ij}$, which must be modelled.  In this paper, we use the popular \komegasst{} model as a baseline model.  The model defines transport equations for the turbulence kinetic energy $k := \frac 1 2 \tau_{ii}$, and the specific turbulence dissipation rate $\omega$:
\begin{align}  \label{kw SST}
U_j \frac{\partial k}{\partial x_j} &= \underbrace{\tau_{ij} \frac{\partial U_i}{\partial x_j}}_{P_k} - \beta^* k\omega + \frac{\partial}{\partial x_j} \left[(\nu + \sigma_k\nu_t) \frac{\partial k}{\partial x_j} \right], \\
  \label{eq:omega}
U_j \frac{\partial \omega}{\partial x_j} &= \frac{\gamma}{\nu_t}\tau_{ij} \frac{\partial U_i}{\partial x_j} - \beta\omega^2 + \frac{\partial}{\partial x_j} \left[(\nu + \sigma_\omega\nu_t) \frac{\partial \omega}{\partial x_j} \right] + 2(1-F_1)\sigma_{\omega2}\frac{1}{\omega}\frac{\partial k}{\partial x_j}\frac{\partial \omega}{\partial x_j} \ .
\end{align}
The eddy viscosity is modelled as
\[
\nu_t := \frac{a_1k}{\max(a_1\omega, SF_2)} \ , 
\]
and the anisotropic part of Reynolds-stress tensor is modelled by the Boussinesq assumption
\[
a_{ij} := \tau_{ij} - \frac 2 3 \delta_{ij} k \simeq  a^B_{ij} := - 2\nu_t S_{ij},\quad \text{where}\: S_{ij}:=\frac{1}{2}\left(\frac{\partial U_i}{\partial x_j} + \frac{\partial U_j}{\partial x_i}\right),
\]
and the isotropic part is absorbed into the pressure.  All remaining terms and coefficients are omitted for brevity, see \citep{Menter1994} for details.
%

The computational mesh used for RANS simulation of the periodic-hill is two-dimensional, with $120 \times 130$ cells in stream-wise and wall-normal directions respectively.  Periodic boundary conditions are set at inlet and outlet, and no-slip conditions on the walls. Volume forcing is used to drive the flow, with a PID controller used to achieve the target velocity.  The simulation is performed at $\mathrm{Re_h} = 10595$ based on the hill height $h$ using second-order accurate SIMPLE solver.  The pressure is solved using geometric-algebraic multigrid with a Gauss-Seidel smoother, while $U_i$, $k$ and $\omega$ are solved using a Diagonal-Incomplete-LU-preconditioned BiCG method.  

\subsection{Large-eddy simulation for the periodic-hills}

The high-fidelity and ``ground-truth'' model in this work is wall-modelled LES with our in-house finite-volume solver INCA.  We use the incompressible staggered-grid version of the solver with a block-Cartesian background mesh and the conservative immersed boundary method of \citet{Meyer2010} for representing the geometry. Our LES follow a holistic modelling approach, where the numerical discretization and the subgrid-scale (SGS) turbulence closure are fully merged: The adaptive local deconvolution method (ALDM) of \citet{Hickel2006,Hickel2007} is a nonlinear finite-volume discretization scheme tailored for LES of turbulent flows. Optimum model form and discretization parameters were learnt by a physics-informed genetic algorithm using turbulence data, spectral analysis and constraints from turbulence theory \citep{Hickel2006}. The near-wall turbulence is modelled rather than resolved~\citep{Hickel2012,Chen2014}, dictated by the available computational resources and the Reynolds numbers in our case.  A third-order explicit Runge-Kutta scheme is used for time integration and the pressure-correction Poisson equation is solved using BiCGstab with incomplete LU and algebraic-multigrid as preconditioners. Overall, the our LES solver is very similar to that used by \citet{Hickel2008} for wall-resolved LES of the periodic-hill flow, and largely identical to that used by \citet{Meyer2013} for wall-modelled LES of a low-speed aerospace application.

\begin{figure}
	\centering
	\includegraphics[width=0.85\textwidth]{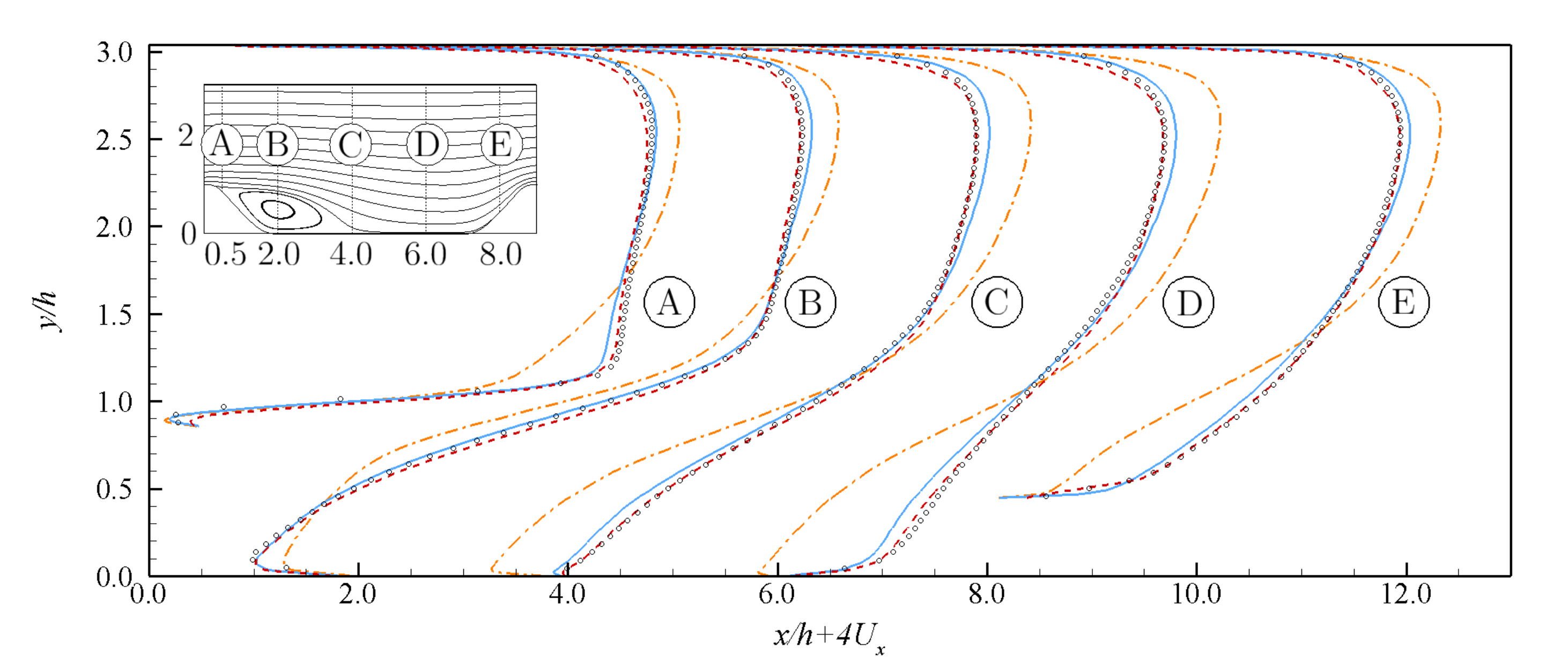}
	\caption{Mean velocity profiles for: reference LES ({\color{CornflowerBlue}{\textemdash}})~\citep{Breuer2009}; PIV ($\circ$)~\citep{Rapp2009}, our LES ({\color{red}- - - -}) and RANS with k-$\omega$ SST turbulence modelling ({\color{YellowOrange}-$\cdot$-$\cdot$-}).}
	\label{fig:ValidationLES}
\end{figure}

A parameter study was performed on the effects of the span-wise extent of the domain, the mesh resolution, and the solution averaging time \citep{Gomez2018}.  A good compromise of accuracy and cost was achieved for a spanwise extent of $L_z = 4.5h$ (with $h = $ hill height), about $1.2$ million cells (giving $y^+ < 20$ everywhere), and averaging over $\simeq 55$ flow-through times started after an initial transient of $\simeq 35$ flow-through times.  With these parameters, converged statistics are obtained in less than 2 days on a dual Intel${}^{(R)}$ Xeon${}^{(R)}$ CPU E5-2670v2. Figure~\ref{fig:ValidationLES} shows our LES results for the mean velocity profiles at several streamwise locations for the standard periodic-hill geometry and the well-resolved LES data of \citet{Breuer2009} ($13.1$ million cells) and the reference experiments of \citet{Rapp2009}. We observe that our INCA LES results are in very good agreement with the reference data. 

Also shown in Figure~\ref{fig:ValidationLES} are the RANS results for the baseline \komegasst{} model described in Section~\ref{s:rans}.  The RANS simulation significantly over-predicts the length of the separated flow region, which  distorts the flow and leads to large errors everywhere in the domain. Having flow-features very dissimilar to the high-fidelity LES, the baseline RANS is therefore likely inappropriate as a low-fidelity model in a bi-fidelity optimization or surrogate modelling procedure.

\section{Enhancing RANS with data-driven turbulence modelling}

In this section we briefly describe the SpaRTA procedure to generate improved RANS models from LES data -- see \cite{Schmelzer2020} for a detailed discussion.  The procedure has two main stages:
\begin{enumerate}
\item We use the \frozen{} approach \citep{Schmelzer2020} to solve for corrective fields for the anisotropy tensor, and turbulence production.  These fields, when directly injected into the RANS solver, lead the solver to reproduce the LES mean-flow.
\item We use deterministic symbolic regression to find a low-complexity approximate map from the resolved mean-flow quantities to these corrective fields.  This map represents our correction to the baseline turbulence model.
\end{enumerate}
The result is a RANS closure model customized for a particular flow.
  
\subsection{Identifying model-form error with the \frozen{} approach}
\label{s:frozen}

It is well known \citep{Xiao2019} that injecting LES/DNS Reynolds stresses directly into RANS solvers does not always improve accuracy of the mean-flow prediction.  This issue can be partially addressed by using a field inversion procedure to identify corrective fields that result in the correct mean-flow \citep{Duraisamy2019}.  In our problem full-field LES data is available, and so a costly inversion procedure is unnecessary.  Instead we define an iteration that leads to equivalent corrective fields.

\begin{figure}
	\centering
	\begin{subfigure}[t]{0.9\textwidth}
		\includegraphics[width=0.95\textwidth]{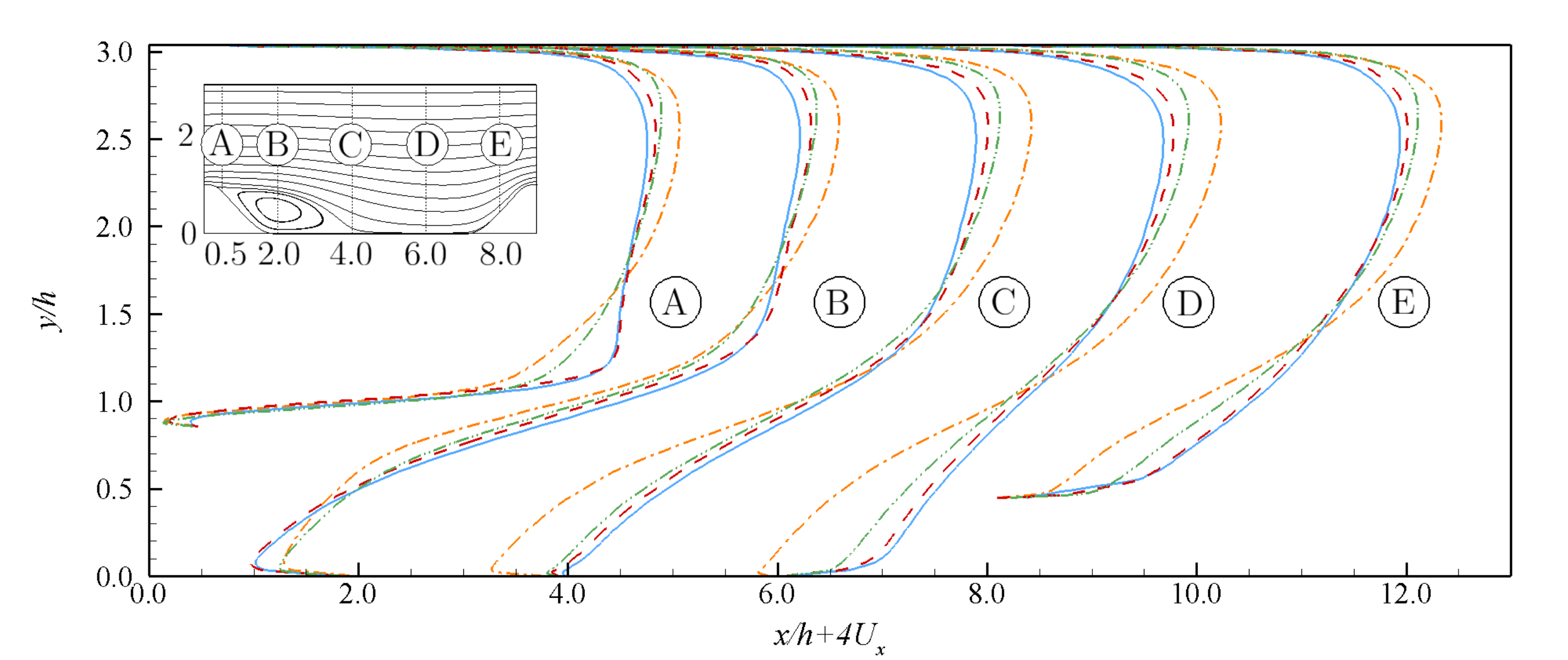}
		\subcaption{PH case ($\psi=1.0$)}
		\label{fig:PH1.0}
	\end{subfigure}%
    \\
	\begin{subfigure}[t]{0.9\textwidth}
        \includegraphics[width=0.95\textwidth]{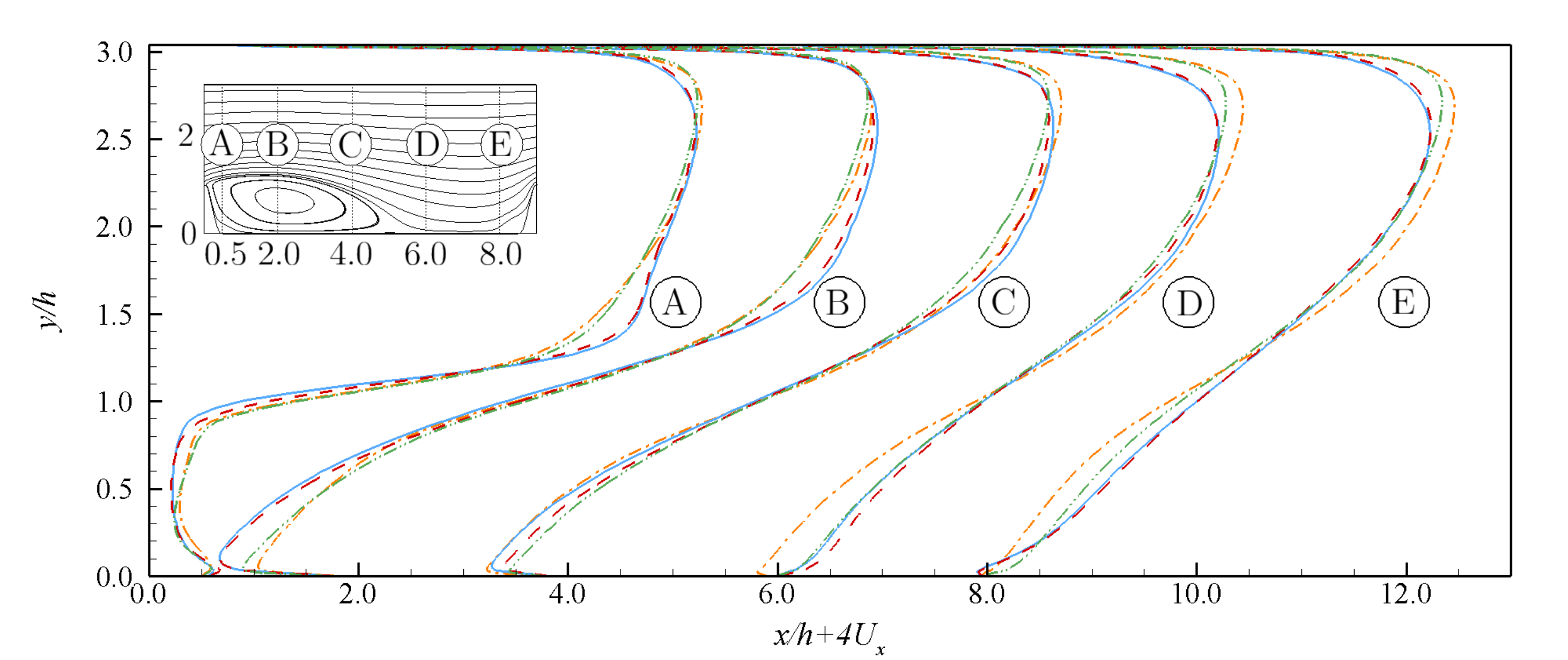}
        \subcaption{PH case ($\psi=0.25$)}
        \label{fig:PH0.25}
	\end{subfigure}
        \\
    \begin{subfigure}[t]{0.9\textwidth}
    	\includegraphics[width=0.95\textwidth]{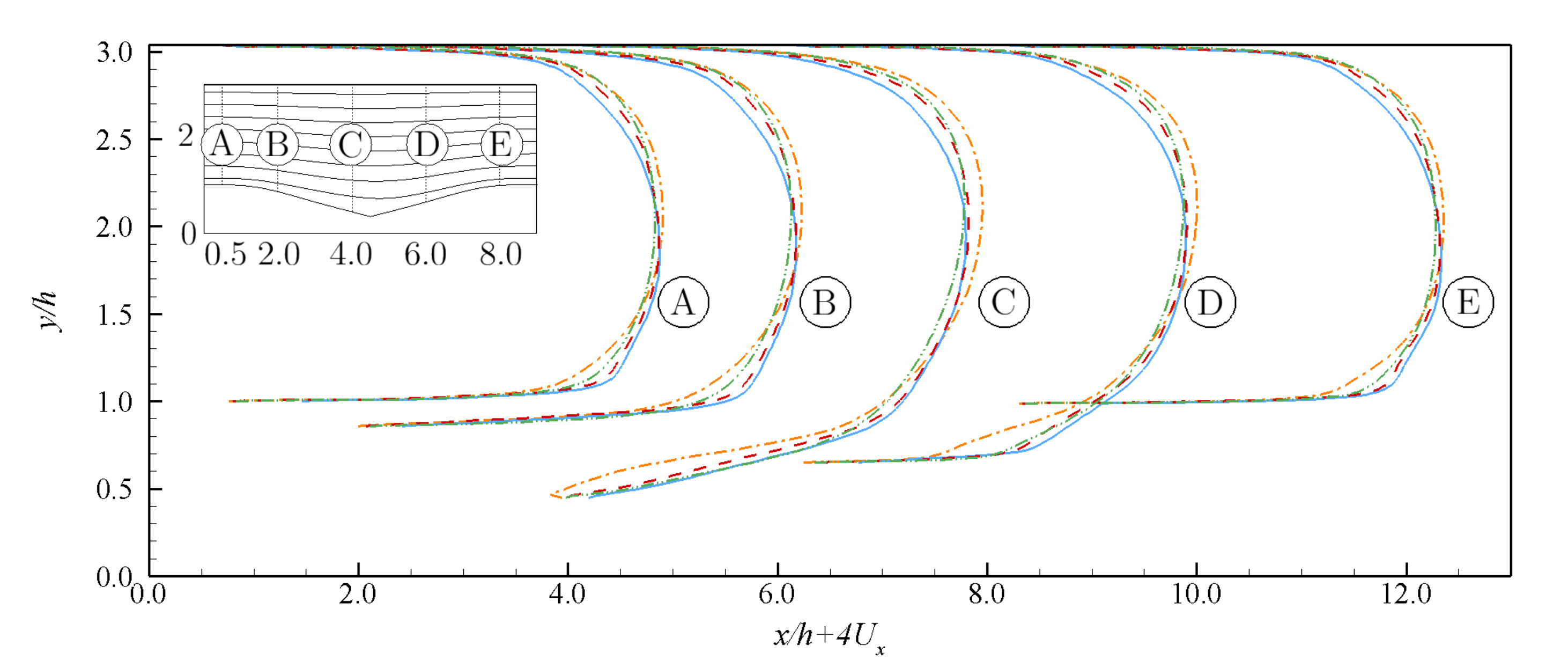}
    	\subcaption{PH case ($\psi=4.0$)}
    	\label{fig:PH4.0}
    \end{subfigure}
	\caption{Mean velocity profiles for PH cases: reference LES ({\color{CornflowerBlue}{\textemdash}})(INCA), Frozen RANS ({\color{red}- - - -}), SpaRTA RANS ({\color{green}-$\cdot\cdot$-$\cdot\cdot$-}) and RANS with k-$\omega$ SST turbulence modelling ({\color{YellowOrange}-$\cdot$-$\cdot$-}).}
	\label{fig:PHcases}
\end{figure}

Let LES statistics be denoted with a superscript ``$\star$'', so $b_{ij}^\star = a_{ij}^\star / 2k^\star$ is the normalized LES anisotropy tensor, $k^\star$ is the LES turbulence kinetic energy, and $U^\star$ the LES mean velocity field.  We define a correction to the normalized anisotropy tensor $b_{ij}^{\Delta}$ via:
\begin{equation}
  b_{ij}^\star = b_{ij}^{\star B} + b_{ij}^{\Delta}, \quad b_{ij}^{\star B} := -\frac{\hat\nu_t}{k^\star} S_{ij}^\star,
  \label{delta bij}
\end{equation}
where the correction is defined with respect to the Boussinesq assumption evaluated using the LES strain-rate tensor $S_{ij}^\star$, $k^\star$ and eddy-viscosity $\hat\nu_t$. \citet{Parneix1998} and \citet{Weatheritt2017} model the specific dissipation rate $\omega$ by passively evaluating the $\omega$-equation \eqref{eq:omega}, using LES data for other quantities. We further extend this procedure by requiring that the $k$ equation also satisfies the LES data, the so-called \frozen{} approach. To this end we introduce a (spatially varying) correction $\hat{\textbf{R}}$ to the $k$- and $\omega$-equations giving:
\begin{align}
  \label{eq:kcorr}
U_j^\star \frac{\partial k^\star}{\partial x_j} &= P_k^\star + \hat{\textbf{R}} - \beta^* k^\star\hat{\omega} + \frac{\partial}{\partial x_j} \left[(\nu + \sigma_k\hat\nu_t) \frac{\partial k^\star}{\partial x_j} \right], \\
U_j^\star \frac{\partial \hat{\omega}}{\partial x_j} &= \frac{\gamma}{\hat\nu_t}\left(P_k^\star + \hat{\textbf{R}} \right) - \beta\hat{\omega}^2 + \frac{\partial}{\partial x_j} \left[(\nu + \sigma_\omega\hat\nu_t) \frac{\partial \hat{\omega}}{\partial x_j} \right] + 2(1-F_1)\sigma_{\omega2}\frac{1}{\hat\omega}\frac{\partial k^\star}{\partial x_j}\frac{\partial \hat{\omega}}{\partial x_j},
\end{align}
in which quantities with a ``\textasciicircum'' are to be solved for, and remaining quantities derive from the LES data.  These equations are solved readily in a loosely coupled fashion, and the resulting $\hat\omega$ is used to compute $b_{ij}^{\Delta}$ from \eqref{delta bij}.  Thus two corrective fields result, the tensor-field $b_{ij}^{\Delta}$ and the scalar-field $\hat{\textbf{R}}$.  When these fields are injected into the RANS solver, the resulting mean-fields consistently agree well with LES, see Figure~\ref{fig:PHcases}.  This is true for both the baseline geometry $\psi = 1$, and steeper and shallower hills with $0.25 \leq \psi \leq 4$ - i.e.\ the full design space.
%

\subsection{Deterministic symbolic regression for corrective field modelling}
\label{s:sparta}

Having identified corrective fields, to make predictions it is necessary to model these fields in terms of known (mean-flow) quantities. Following \citet{Pope1975}, we assume that the non-dimensional strain-rate and rotation-rate tensors 
\[
\tilde S_{ij} := \frac{1}{\omega}S_{ij},\qquad
\tilde \Omega_{ij} := \frac{1}{2\omega} \left(\frac{\partial U_i}{\partial x_j} - \frac{\partial U_j}{\partial x_i} \right)
\]
are sufficient to describe the corrective fields, leading to \citeauthor{Pope1975}'s well-known basis-tensor series. The Cayley-Hamilton theorem dictates that the most general form of the function $b^\Delta_{ij}(\tilde{S}_{ij}, \tilde{\Omega}_{ij})$ is (under the assumptions of analyticity and Galilean invariance):
\begin{align}
\label{CHtheroem}
b^\Delta_{ij} (\tilde{S}_{ij}, \tilde{\Omega}_{ij}) = \sum_{n=1}^{10} T_{ij}^{(n)} \alpha_n(\lambda_1,\dots,\lambda_5),
\end{align}
where $T_{ij}^{(n)}$ are ten basis tensors, $\lambda_m$ are the five invariants of $\tilde S$ and $\tilde \Omega$, and $\alpha_n(\cdot)$ are arbitrary scalar functions of these invariants.  In this paper, we only consider the first three base tensors, and first two invariants, which are given by:
\begin{align*}
T_{ij}^{(1)} &= \tilde{S}_{ij} & \lambda_1 &= \tilde{S}_{mn}\tilde{S}_{nm}, \\
T_{ij}^{(2)} &= \tilde{S}_{ik}\tilde{\Omega}_{kj} - \tilde{\Omega}_{ik}\tilde{S}_{kj} & \lambda_2 &= \tilde{\Omega}_{mn}\tilde{\Omega}_{nm}, \\
T_{ij}^{(3)} &= \tilde{S}_{ik}\tilde{S}_{kj} - \frac{1}{3} \delta_{ij} \tilde{S}_{mn}\tilde{S}_{nm}.
\end{align*}
To model $\hat{\textbf{R}}$ we assume that it takes the form
\[
\hat{\textbf{R}} = 2 k \frac{\partial U_i}{\partial x_j} b^R_{ij},
\]
for some tensor-field $b^R_{ij}$, modelled similarly to $b^\Delta_{ij}$ by \eqref{CHtheroem}.  It remains to fit a function of the form \eqref{CHtheroem} to the LES data, for which we use symbolic regression.

For details refer to \cite{Schmelzer2020}.  In short, we form a large {\it library} of candidate basis functions, consisting of powers, and products of the invariants $\lambda_m$.  We then regress the data -- consisting of $(\lambda_m, T^{(n)}_{ij}, b^\Delta_{ij})$ triples at each spatial mesh point -- against this library with sparsity-promoting $\ell^1$ regularization.  This selects a small set of basis functions that represent the data well.  Concretely, let the regressing function be:
\[
b^\Delta_{ij} \simeq \sum_{n=1}^{4} T_{ij}^{(n)} \left\{\sum_{m=1}^{M} \phi_m(\lambdab)\cdot\theta_n^m\right\} = L(\Tb, \lambdab; \thetab),
\]
where $\boldsymbol{\theta}$ are coefficients of the $M$ basis functions in the library $\phi_m(\cdot)$ for each basis tensor, and $L(\cdot)$ is the implied operator, which is linear in $\thetab$.  Then we solve the elastic-net regularized regression problem:
\begin{align}
  \label{ModelSelection}
\mathbf{\thetab}_\mathrm{fit} := \operatorname*{arg\,max}_{\thetab \in \Theta} \left\| L(\Tb^\star, \lambdab^\star; \thetab) - b^\Delta_{ij}\right\|_2^2 + \sigma\rho \left\| \thetab\right\|_1 + \sigma(1-\rho) \left\| \thetab\right\|_2^2,
\end{align}
where the first norm is over all mesh points of all training cases, and once again $\star$ denotes quantities evaluated from the LES data.  We solve for a number of different values of $\sigma$ and $\rho$, to obtain a number of candidate sparsity patterns, and we discard the values of the nonzero coefficients $\thetab$.  For each of the sparsity patterns, we perform ridge regression, using only the basis functions identified in the sparse regression.  Let $\thetab^s$ be a vector of non-zero coefficients identified by \eqref{ModelSelection}, then we solve
\begin{align}
\mathbf{\thetab^s}_\mathrm{fit} := \operatorname*{arg\,max}_{\thetab \in \Theta} \left\| L(\Tb^\star, \lambdab^\star; \thetab^s) - b^\Delta_{ij}\right\|_2^2 +  \sigma_r \left\| \thetab^s\right\|_2^2,
\end{align}
for each sparsity pattern.  The final model is selected based on a compromise between sparsity, goodness-of-fit, and ideally performance in cross-validation if data is available \citep{Schmelzer2020}.  This two step procedure allows us to independently control the level of sparsity (with $\rho$, $\sigma$), and the magnitude of the resulting coefficients (with $\sigma_r$), which can otherwise become very large.

Applying this procedure to our LES data for the standard P-H case ($\psi = 1$), the resulting algebraic models for $b_{ij}^{\Delta}$ and $\hat{\textbf{R}}$ are:
\begin{align} \label{modelbs}
b^\Delta_{ij} \simeq M_{b_{ij}^{\Delta}} =  2.8 T_{ij}^{(2)}; \quad
\hat{\textbf{R}} \simeq M_{R} = 0.4 T_{ij}^{(1)}\frac{\partial U_i}{\partial x_j}.
\end{align}

\begin{figure}
	\centering
	\begin{subfigure}[t]{0.4\textwidth}
		\centering
		\includegraphics[width=0.99\textwidth]{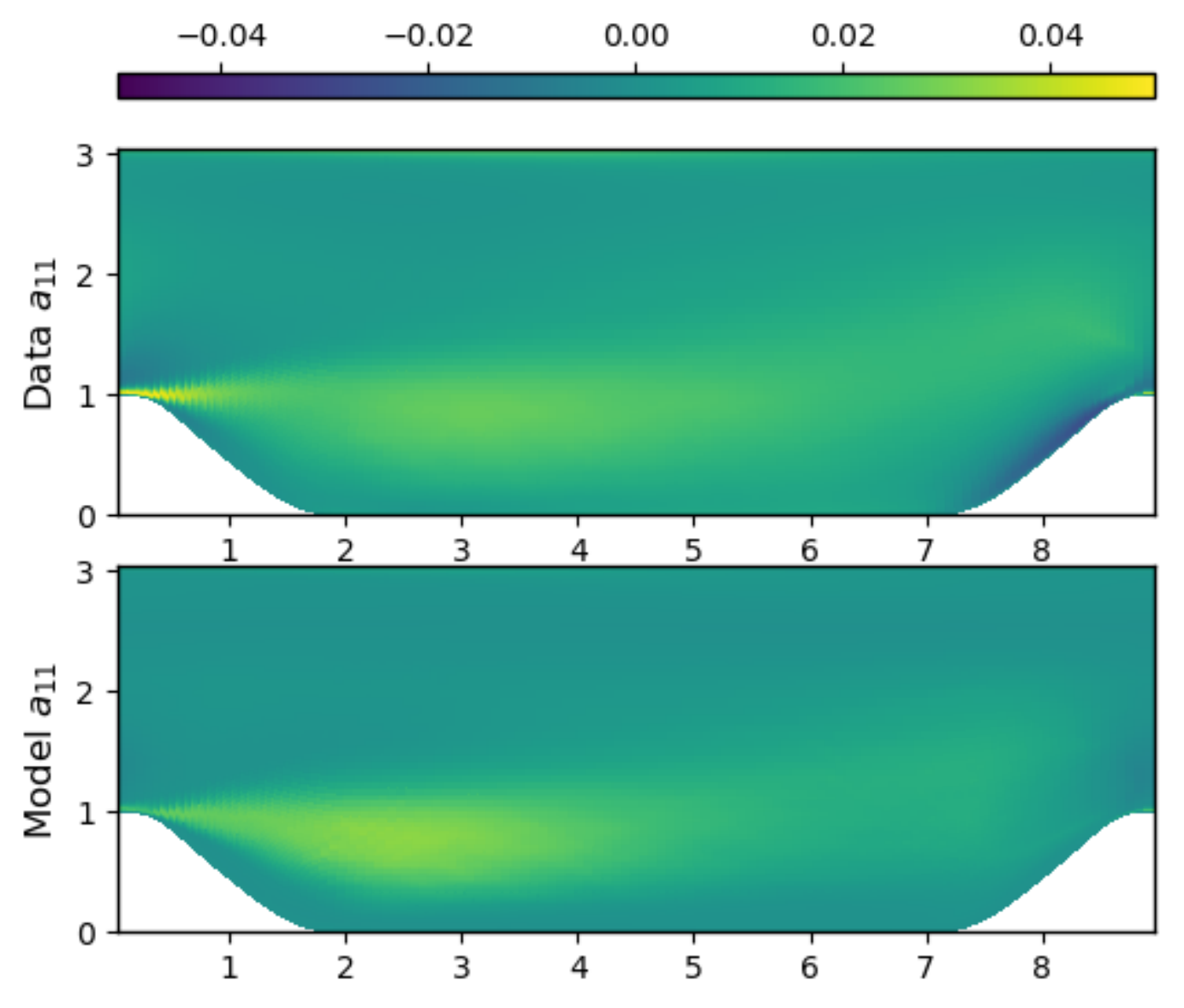}
	\end{subfigure}%
	\begin{subfigure}[t]{0.4\textwidth}
		\centering
		\includegraphics[width=0.99\textwidth]{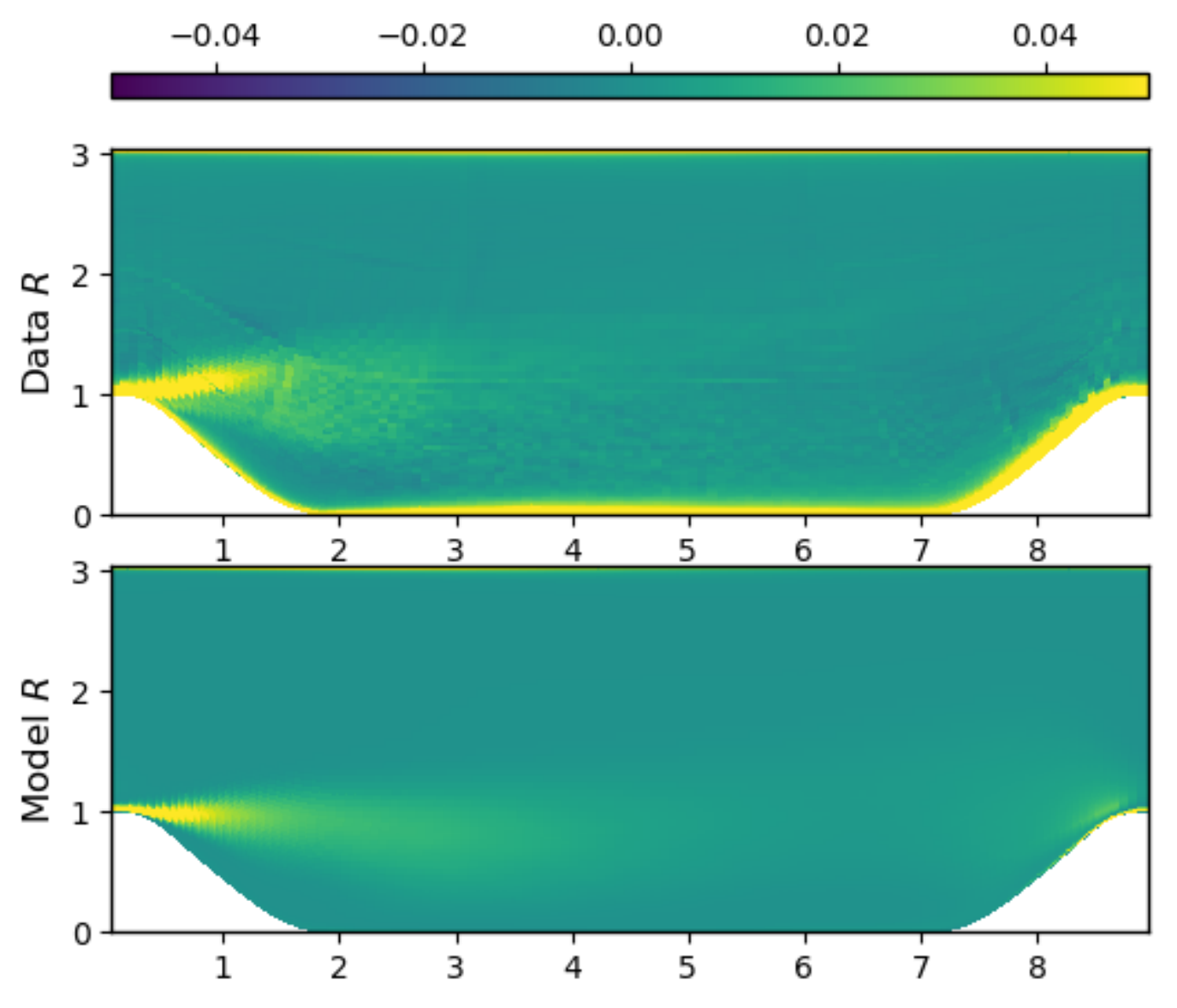}
	\end{subfigure}
	\caption{Comparison of LES and model prediction (left: $a_{11}$ in Reynolds stress; Right: $R$ residual of k transport equation)}
	\label{fig:aijR}
\end{figure}

Figure~\ref{fig:aijR} shows the optimal $a_{ij} = 2kb^\Delta_{ij}$ and $\hat{\textbf{R}}$ identified for the P-H case using the $k$-frozen approach of the previous section.  The predictions of the model \eqref{modelbs} are also shown.  The agreement is generally good, with the notable exception of the large $\hat{\textbf{R}}$ near the wall, which is not reproduced by \eqref{modelbs}.  This failure can be attributed to the non-dimensionalization of $S$, $\Omega$ by the specific dissipation rate $\omega$, which becomes large close to the wall, pulling all basis-tensors $T^{(n)}$ down to zero.  This effect can be overcome by using coefficients $\alpha_n$ that go to infinity near the wall, as in \citep{Kaandorp_2020}.  However, we consider this unnecessary, as corrections close to the wall, especially in $b^\Delta$, have very little effect on the mean-flow.

Implementing $M_{b_{ij}^{\Delta}}$ and $\hat{\textbf{R}}$ in the RANS solver, we verify that the mean-flow of the training case is reproduced, and we make predictions for $\psi=0.25$ and $\psi=4.0$ and compare with our LES reference, see Figure~\ref{fig:PHcases}.  In all cases the mean-flow is significantly improved over the baseline \komegasst{}, but do not always reach the accuracy of the directly injected corrective fields shown ("frozen RANS" in Figure~\ref{fig:PHcases}) which represent a best case scenario.

\section{Bi-fidelity optimization loop with custom RANS as low-fidelity}
%
We aim to efficiently solve the discretized PDE-constrained minimization problem
\[
\min_{\psi\in\Psi} J(\psi; U)\quad\text{subject to} \quad R_\mathrm{LES}(\psi; U) = 0,
\]
where $\psi$ are geometric design-variables in the design space $\Psi$, $U$ is the full flow-state, $J$ is the cost-function and $R_\mathrm{LES}$ is the (highest-fidelity) discretized PDE operator (which depends on $\psi$ via boundary-conditions).  We assume that no significant modelling is required to evaluate $J$ given $U$, which is the case for most cost-functions of interest.  We also have available a low-fidelity discretized operator $R_\mathrm{RANS}[M](\psi; U)$, which depends on some closure model $M$ derived using the procedure of the previous section.

\begin{figure}
	\centering
	\includegraphics[width=0.7\textwidth]{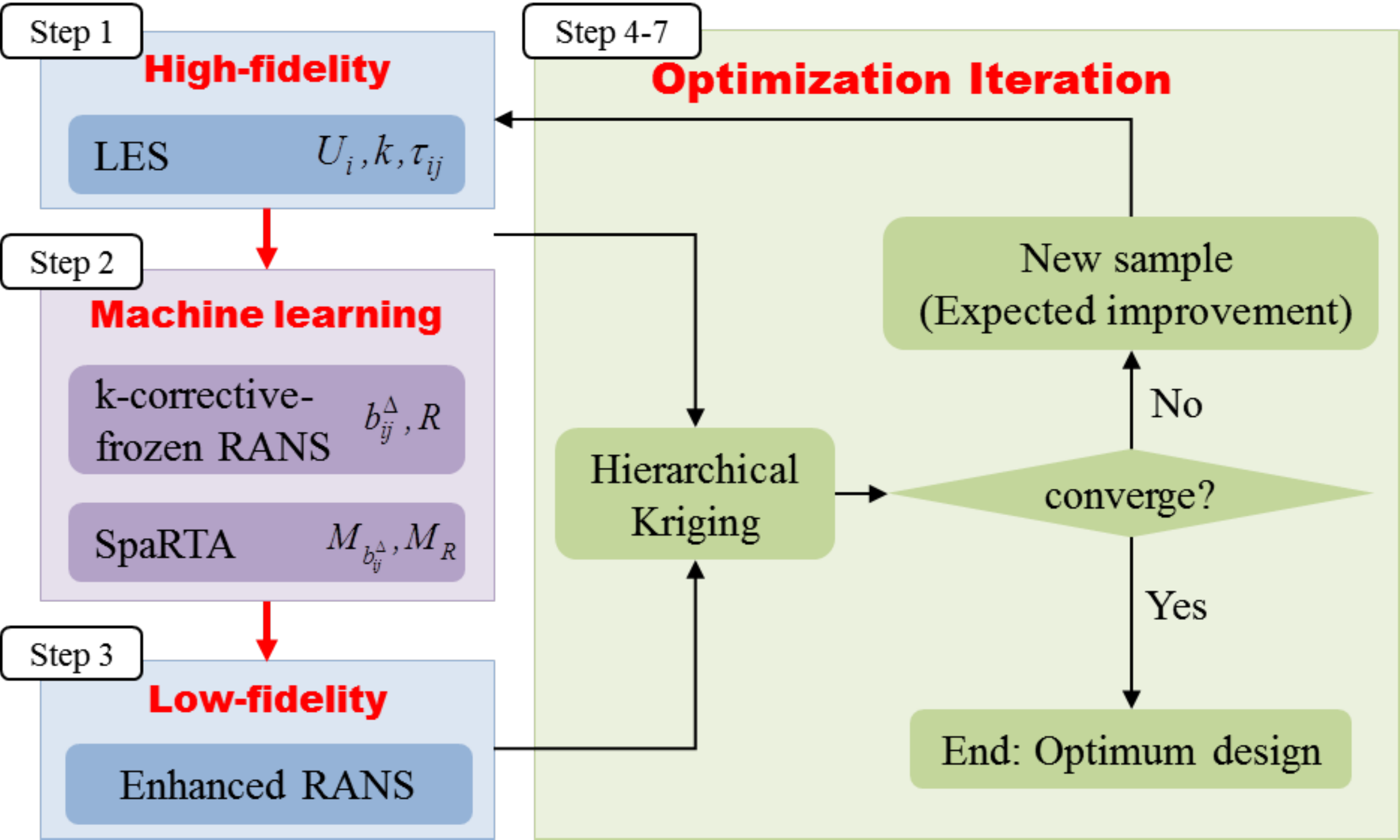}
	\caption{Flowchart of the adaptive variable-fidelity optimization}
	\label{fig:AVFO}
\end{figure}

The outline of our proposed bi-fidelity optimization is given in Figure~\ref{fig:AVFO}.  The steps are:
\begin{enumerate}
\item Solve $R_\mathrm{LES}(\psi_0, U)=0$ for the baseline geometry $\psi=\psi_0$, to obtain turbulence statistics $U^\star$, $\tau^\star_{ij}$, $k^\star$.  Let $i=0$.
\item Identify the model-form error terms $b_{ij}^{\Delta}$ and $\hat{\textbf{R}}$ using the \frozen{} approach of Section~\ref{s:frozen} and the LES data from Step 1. Subsequently find the algebraic models $M_{b_{ij}^{\Delta}}^0$, $M_{R}^0$ using Section~\ref{s:sparta}, to obtain $R_\mathrm{RANS}[M^0](\psi; U)$.
\item Generate $N \gg 1$ low-fidelity sample points $\psi'_1,\dots,\psi'_N$ by design of experiment (DoE) (e.g.\ Latin-hypercube sampling).  Solve $R_\mathrm{RANS}[M^i](\psi'_j; U) = 0$ and evaluate $J$ at each sample $j\in\{1,\dots,N\}$.
\item Build a bi-fidelity Kriging surrogate of $J$ using LES samples at $\psi_1,\dots,\psi_i$, and low-fidelity samples at $\psi'_1,\dots,\psi'_N$.
\item Choose new high-fidelity sample $\psi_{i+1}$ by maximum expected improvement on the surrogate of Step 4.  Evaluate LES at this point.
\item Train a new low-fidelity model $M^{i+1}$ based on LES data at $\psi_1,\dots,\psi_{i+1}$.
\item $i \leftarrow i+1$.  If convergence is achieved, then terminate; otherwise goto Step 3.
\end{enumerate}

Note that in this work the low-fidelity model depends on the result of the high-fidelity (and the locations at which this is evaluated), in contrast to standard multi-fidelity models.  This presents a challenge from the standpoint of statistical modelling of the bi-fidelity system -- for example in cokriging, where a prior correlation relationship must be specified between the fidelities.  This will be the subject of future work.  We bypass the issue here by using hierarchical Kriging. The optimization algorithm is conducted using an in-house tool "SurroOpt"~\citep{Han2016}.

\subsection{Variable-fidelity surrogate modelling: Hierarchical Kriging}

The hierarchical Kriging model \citep{Han2012b} (HK) is constructed in a recursive way. First, an ordinary Kriging model $\hat{J}_\mathrm{Low}$ is built for the low-fidelity objective function. Then a surrogate for the high-fidelity cost function $\hat{J}$ is built using the scaled low-fidelity Kriging model $\hat{J}_\mathrm{Low}$ as the model trend.  In this way the high-fidelity responses are treated as realizations of a random process:
\begin{equation}
\hat J(\psi) = \rho\hat{J}_\mathrm{Low} + Z(\psi)
\end{equation}
where $Z$ is a stationary Gaussian process with zero mean and a covariance $\mathrm{Cov} [Z(\psi),Z(\psi^{\ast})] = \sigma^2 r(\psi, \psi^{\ast} )$, where $\sigma^2$ is the process variance, and $r(\cdot,\cdot)$ and the correlation function.  In this paper, we use a cubic spline function \citep{Han2012b}, which introduces hyper-parameters $\boldsymbol{\theta}$.

Similar to the standard single-fidelity Kriging model, the corresponding predictor and mean-squared error (MSE) of Kriging for the high-fidelity cost function are
\begin{align} 
\hat{J}(\psi) &= \rho \hat{J}_\mathrm{Low}(\psi) +\boldsymbol{r}^{T}\boldsymbol{R}^{-1}(\boldsymbol{J}-\rho\boldsymbol{F})\\
\varepsilon(\psi)^2 &= \sigma^2\{1-\boldsymbol{r}^T\boldsymbol{R}^{-1}\boldsymbol{r}+[\boldsymbol{r}^T\boldsymbol{R}^{-1}\boldsymbol{F}-\hat{J}_\mathrm{Low}(\psi)]^2/(\boldsymbol{F}^T\boldsymbol{R}^{-1}\boldsymbol{F})^{-1}\}
\end{align}
where
\[
\boldsymbol{F} = [\hat{J}_\mathrm{Low}(\psi_{1}),\dots,\hat{J}_\mathrm{Low}(\psi_{N})] \in \mathbb{R}^{N}.
\]
Here, $\boldsymbol{J}\in\mathbb{R}^N$ is the vector of high-fidelity observations at sample sites $\boldsymbol{\psi} = [\psi_1,\dots,\psi_N]$.  $\boldsymbol{R}$ is the correlation matrix between the observed high-fidelity samples and $\boldsymbol{r}$ is the correlation vector between the observed samples and the predicting point. Therefore, the Kriging model for high fidelity cost function can be regarded as a sum of scaled lower fidelity Kriging predictor and the discrepancy between the scaled lower-fidelity function and higher-fidelity function.

Model fitting of the HK model uses the maximum-likelihood estimate:
\begin{equation}
\max_{\rho, \boldsymbol{\theta}, \sigma^2}  L(\rho, \boldsymbol{\theta}, \sigma^2) = \max_{\rho, \boldsymbol{\theta}, \sigma^2} \frac{1}{\sqrt{2\pi\sigma^2 |\boldsymbol{R}|}}\exp\left[-\frac{1}{2}\frac{(\boldsymbol{J}-\boldsymbol{F}\rho)^{T}\boldsymbol{R}^{-1}(\boldsymbol{J}-\boldsymbol{F}\rho)}{\sigma^2}\right],
\end{equation}
given which an explicit expression for $\rho$ is:
\[
\rho = (\boldsymbol{F}^{T}\boldsymbol{R}^{-1}\boldsymbol{F})^{-1}(\boldsymbol{F}^{T}\boldsymbol{R}^{-1}\boldsymbol{J}).
\]

It remains to specify in what manner new high-fidelity samples are chosen.  We use the maximum expected improvement (EI) principle~\citep{Jones1998,Zhang2018}.  Define the {\it improvement} with respect to a current best value $J_\mathrm{min}$ as (formally):
\[
I(\psi) = \max(J_\mathrm{min} - \hat J(\psi), 0) .
\]
Since the prediction of the HK model at any untried site $\psi$ is the normal distribution $\mathcal{N}(\hat{J}(\psi),\varepsilon(\psi)^2)$, we choose the value of $\psi$ that maximizes the expected improvement $\mathbb{E} I(\psi)$.  This has a closed form expression:
\[
\mathbb{E} I(\psi) = (J_\mathrm{min}-\hat{J}(\psi))\Phi\left(\frac{J_\mathrm{min}-\hat{J}(\psi)}{\varepsilon(\psi)}\right)+\varepsilon(\psi)\phi\left(\frac{J_\mathrm{min}-\hat{J}(\psi)}{\varepsilon(\psi)}\right) ,
\]
where $\phi(\cdot)$ and $\Phi(\cdot)$ are respectively the unit normal density and distribution functions.

\subsection{Periodic hill optimization: Cost function and design-space exploration}

\begin{figure}
	\centering
	\includegraphics[width=0.6\textwidth]{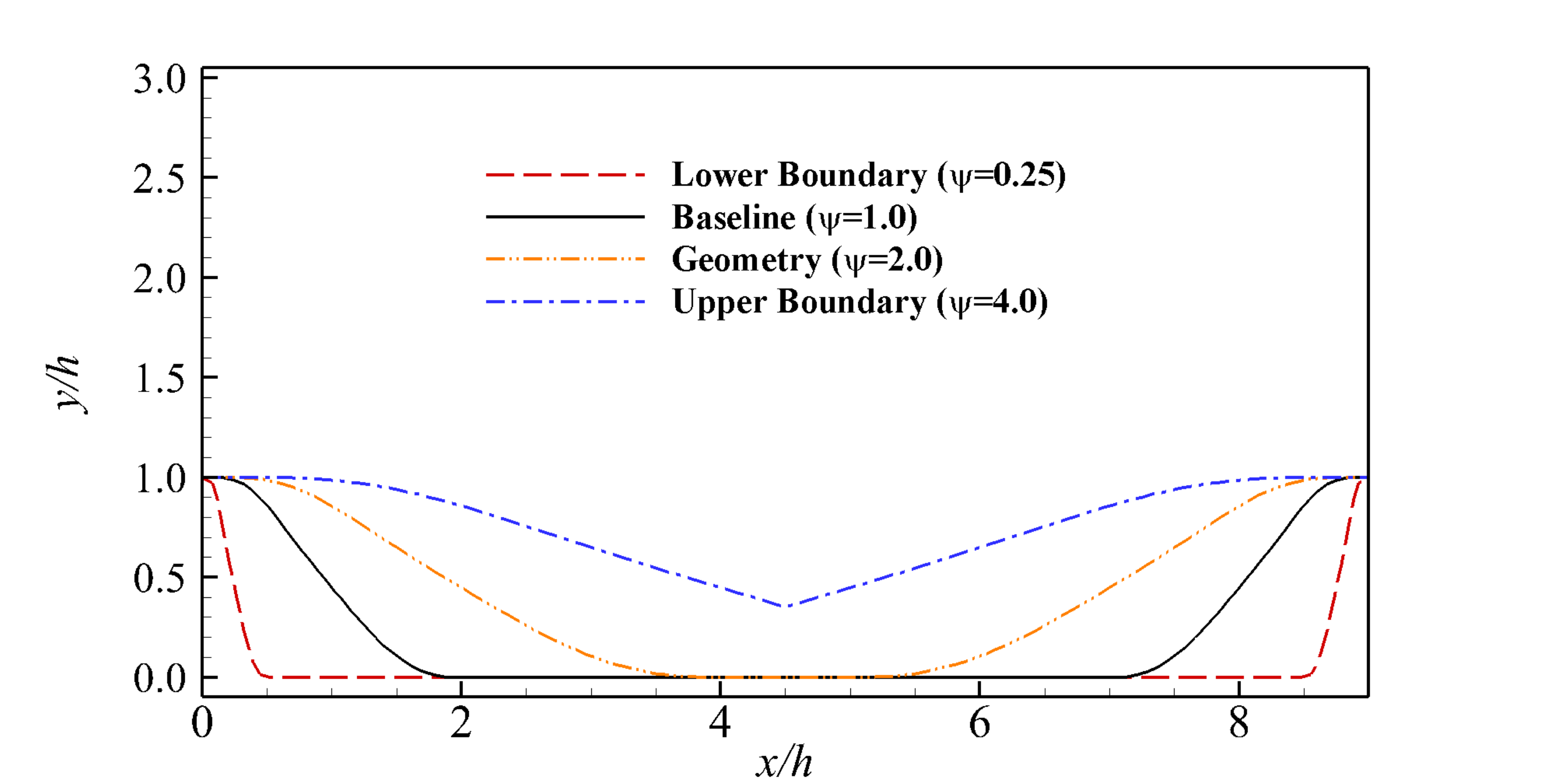}
	\caption{Geometry illustrations for the design space}
	\label{fig:designspace}
\end{figure}

The baseline periodic-hill geometry (hill width $\psi=1$) is the widely known ERCOFTAC Testcase 9.2 with the geometry definition of \citet{Mellen2000}. We choose an objective function depending on the total drag on the hills and averaged turbulence kinetic energy over the whole flow fields at a specified constant mass-flow. To demonstrate optimization in this proof-of-concept study, we choose a cost function has two local optima within the design-space, $\psi \in \Psi = [0.25, 4]$.  The chosen function is: 
\begin{align} \label{obj}
\min_{\psi\in\Psi} \underbrace{-(\bar k-2.5f)/{J_0}}_J, \qquad \Psi = [0.25, 4]
\end{align}
where $\bar k$, is the averaged turbulent kinetic energy over the flowfield, and $f$, the volume forcing term in \eqref{RANS}, acts as a proxy for drag. The cost function $J$ is normalized by the LES result of the baseline geometry, $J_0 := \bar k^\star-2.5f^\star$. The geometries of periodic-hills at the lower- and upper-boundary of the design-space are shown in Figure~\ref{fig:designspace}.

\begin{figure}
	\centering
	\includegraphics[width=0.5\textwidth]{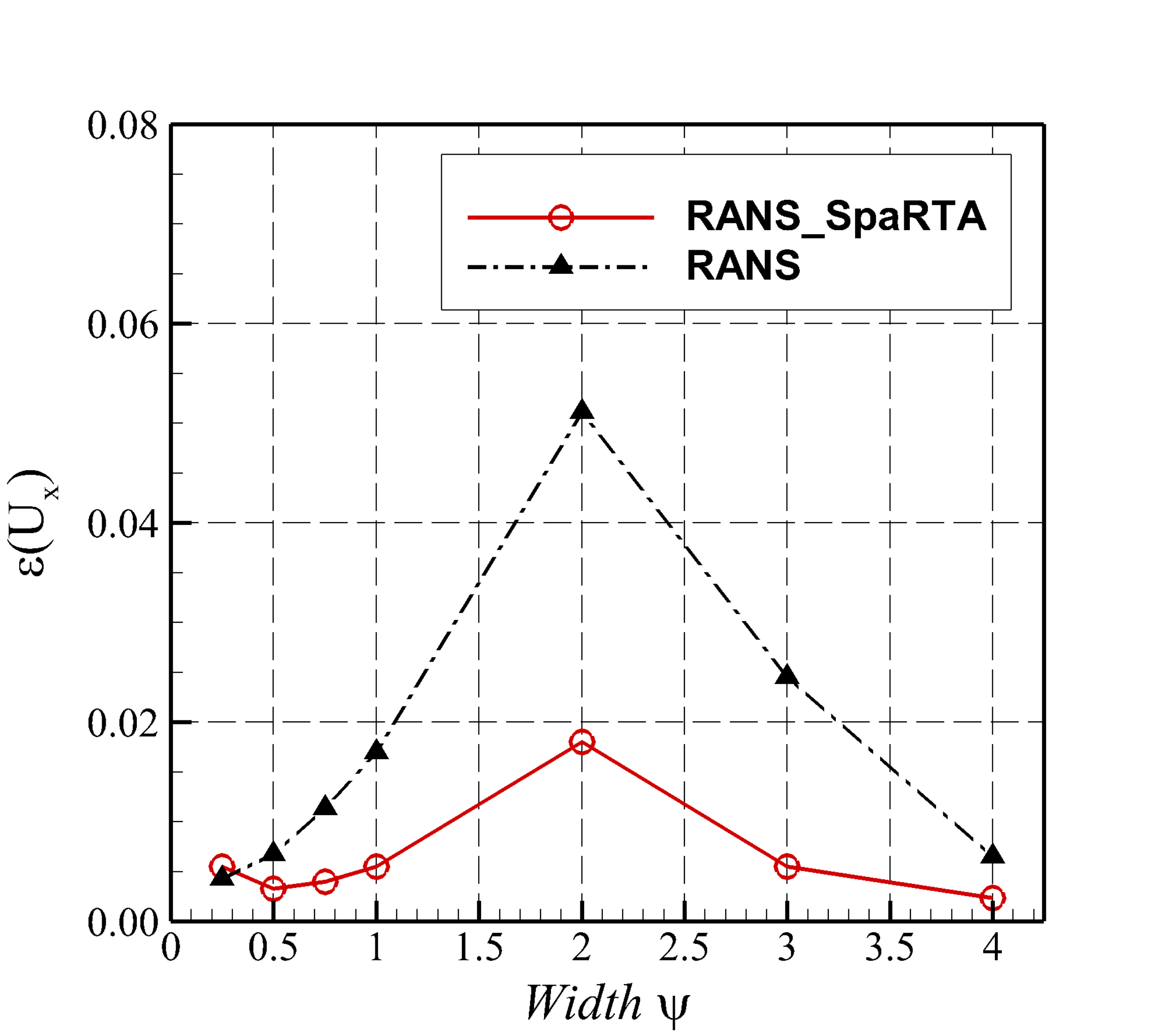}
	\caption{Mean-squared error of mean velocity obtained by custom RANS and baseline RANS}
	\label{fig:mseU}
\end{figure}

\begin{figure}
	\centering
	\begin{subfigure}[t]{0.45\textwidth}
		\centering
		\includegraphics[width=0.99\textwidth]{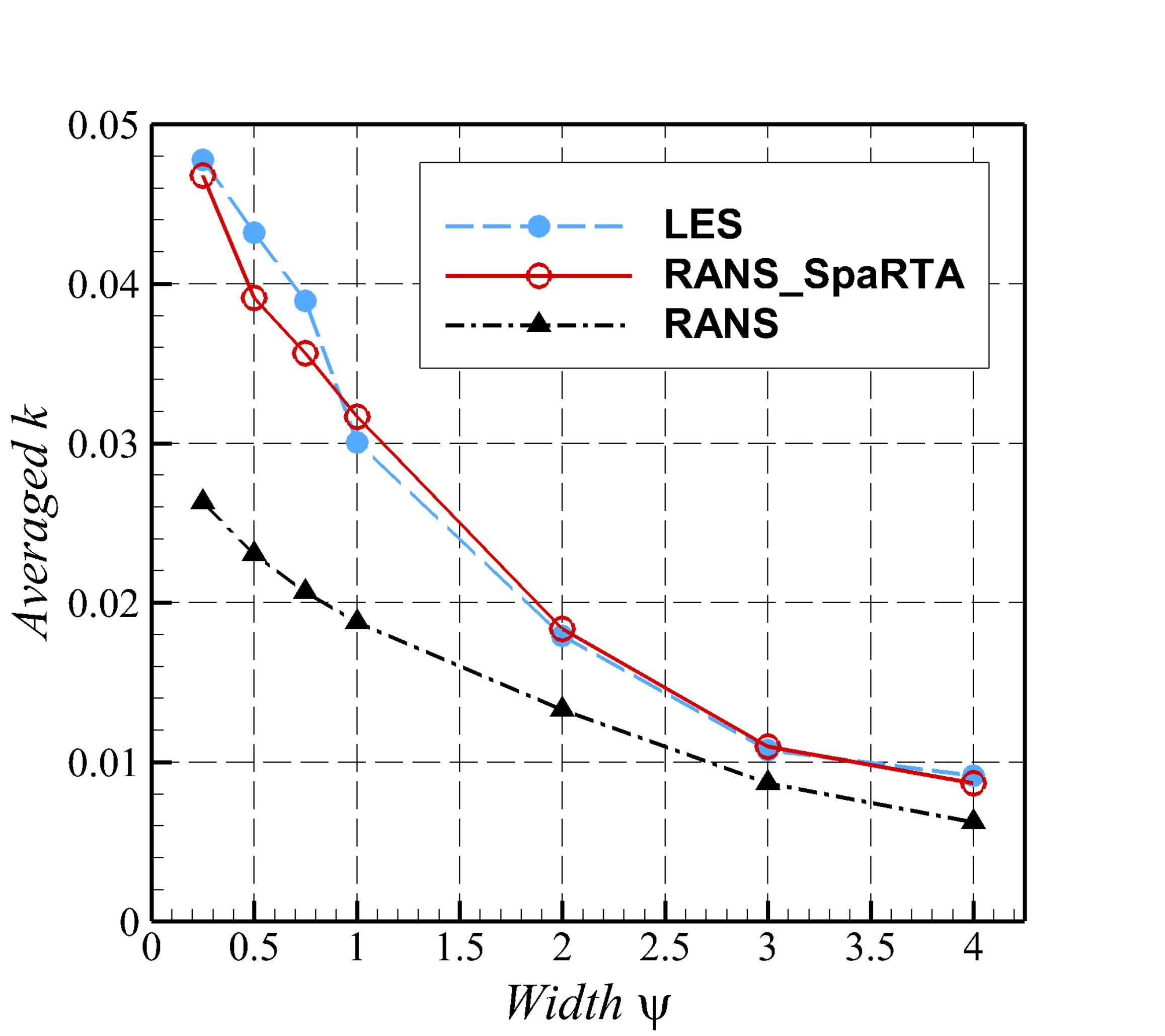}
		\caption{averaged turbulence kinetic energy}
	\end{subfigure}
	\begin{subfigure}[t]{0.45\textwidth}
	    \centering
	    \includegraphics[width=0.99\textwidth]{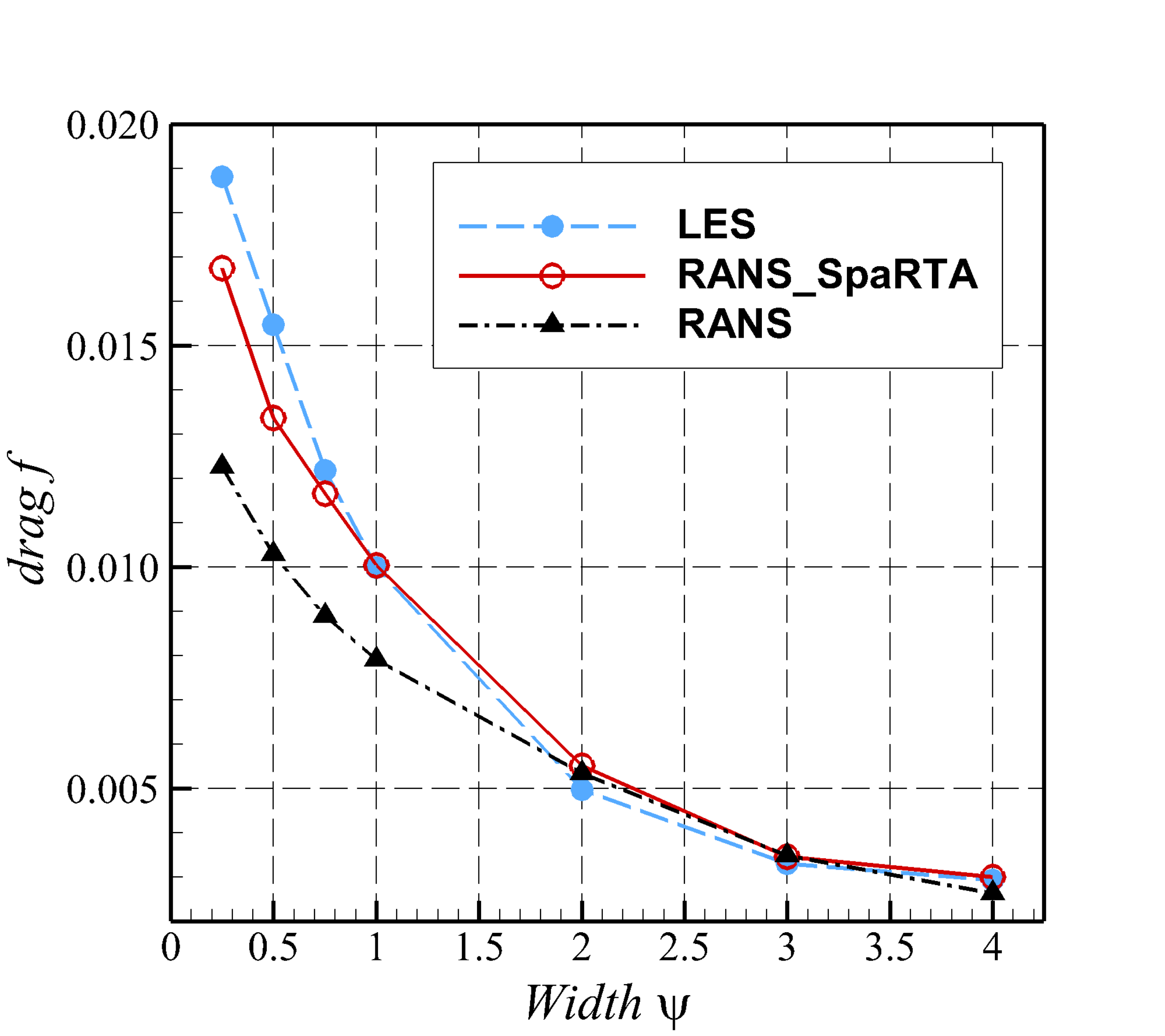}
	    \caption{drag $f$}
    \end{subfigure}%
	\caption{Comparison of cost-function components obtained by LES, custom RANS and baseline RANS.}
	\label{fig:pgAk}
\end{figure}

\begin{figure}
	\centering
	\includegraphics[width=0.5\textwidth]{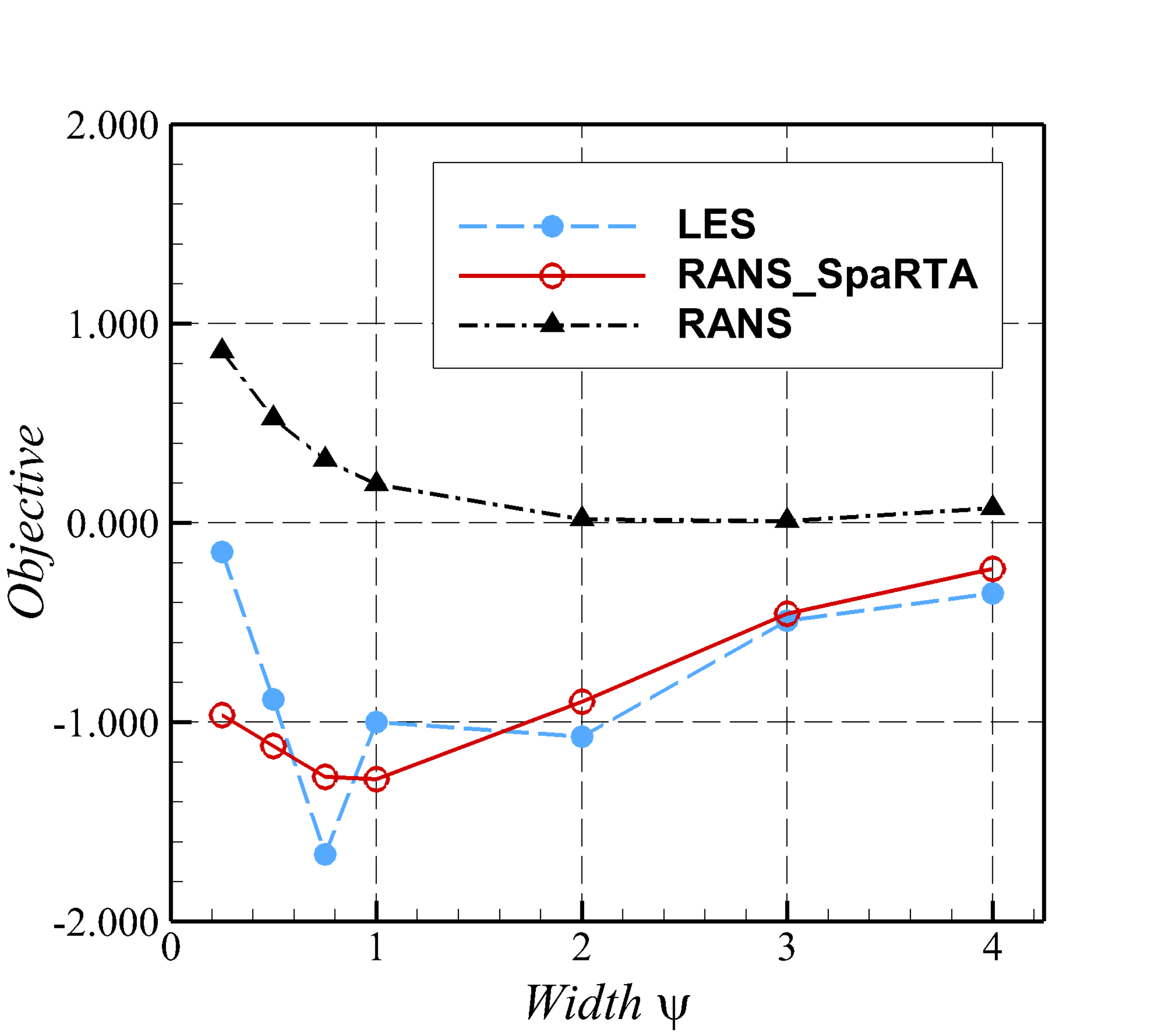}
	\caption{Design-space exploration of the objective function $J$.}
	\label{fig:obj}
\end{figure}

Before the optimization, to evaluate the accuracy of the customized RANS model, we use the model \eqref{modelbs} trained at $\psi=1$ from Section 3.2 to predict the flow at other values of $\psi$, and compare with our LES ground-truth.  In Figure~\ref{fig:mseU}, we see that the mean-velocity prediction is significantly improved compared with the baseline RANS simulation, across the majority of the design-space.  Figure~\ref{fig:pgAk} shows that the prediction of the averaged turbulence kinetic energy $\bar k$ and proxy-drag $f$ are both significantly improved - a consequence of more accurate Reynolds stresses and mean flow fields for all test geometries.  The custom RANS performs best when $\psi \geq 1$, for which the flow shows a medium-sized separation bubble. For the steeper geometries $\psi < 1$, the larger separation bubbles are slightly under-estimated by the custom closure model.  In Figure~\ref{fig:obj}, we observe that the custom RANS matches the main trend of LES very well. In terms of the objective function $J$, the optimization problem has two local minimum points near $\psi = 0.75$ and $\psi = 2.0$, whereas the objective estimated by the custom RANS trained using the baseline data does only show the first one. With additional hi-fidelity LES samples, the custom RANS model will become more accurate during the optimization. The baseline RANS model, however, misses the trend and significantly under-predicts drag and Reynolds stress.

\subsection{Optimization convergence}

\begin{figure}
	\centering
	\begin{subfigure}[t]{0.8\textwidth}
		\includegraphics[width=0.99\textwidth]{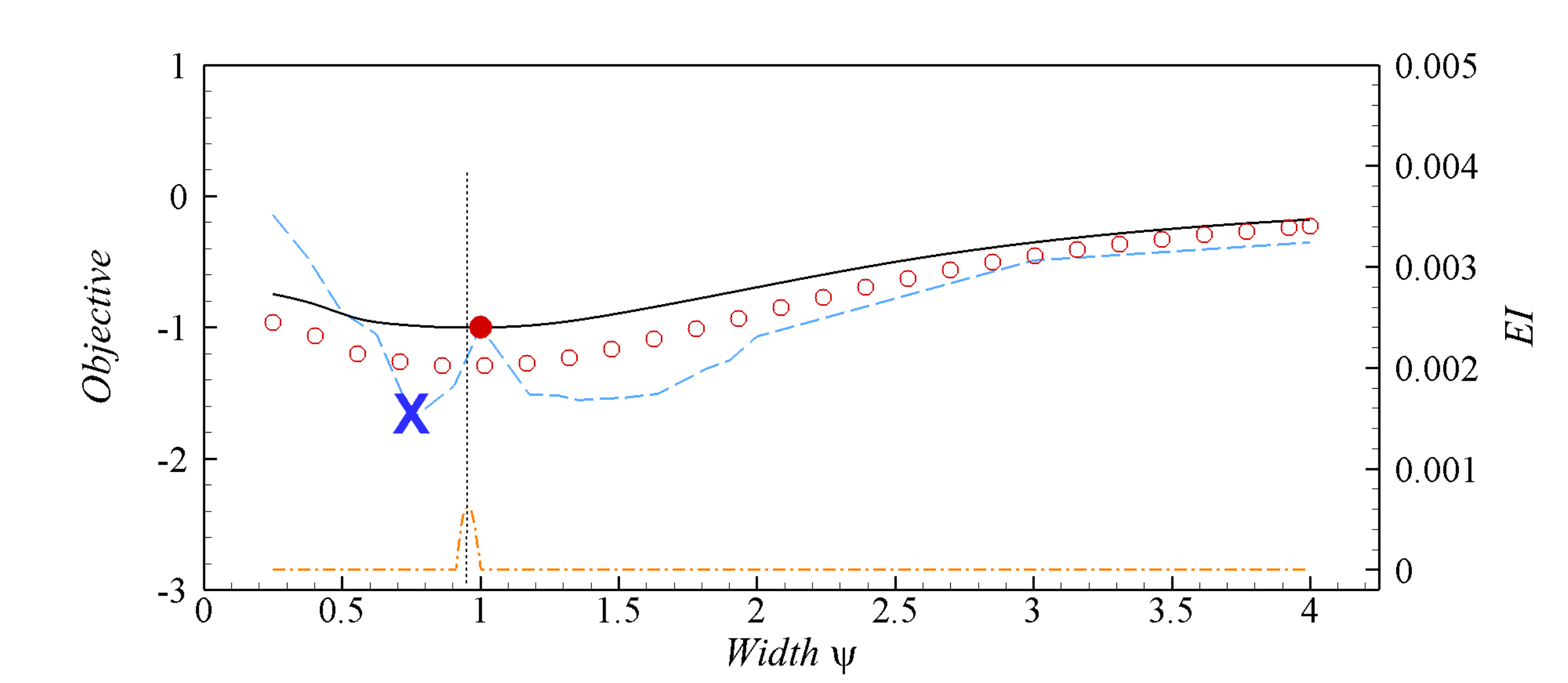}
        \caption{HK surrogate and expect improvement (EI) function at $i=0$}
	\end{subfigure}%
    \\
	\begin{subfigure}[t]{0.8\textwidth}
		\includegraphics[width=0.99\textwidth]{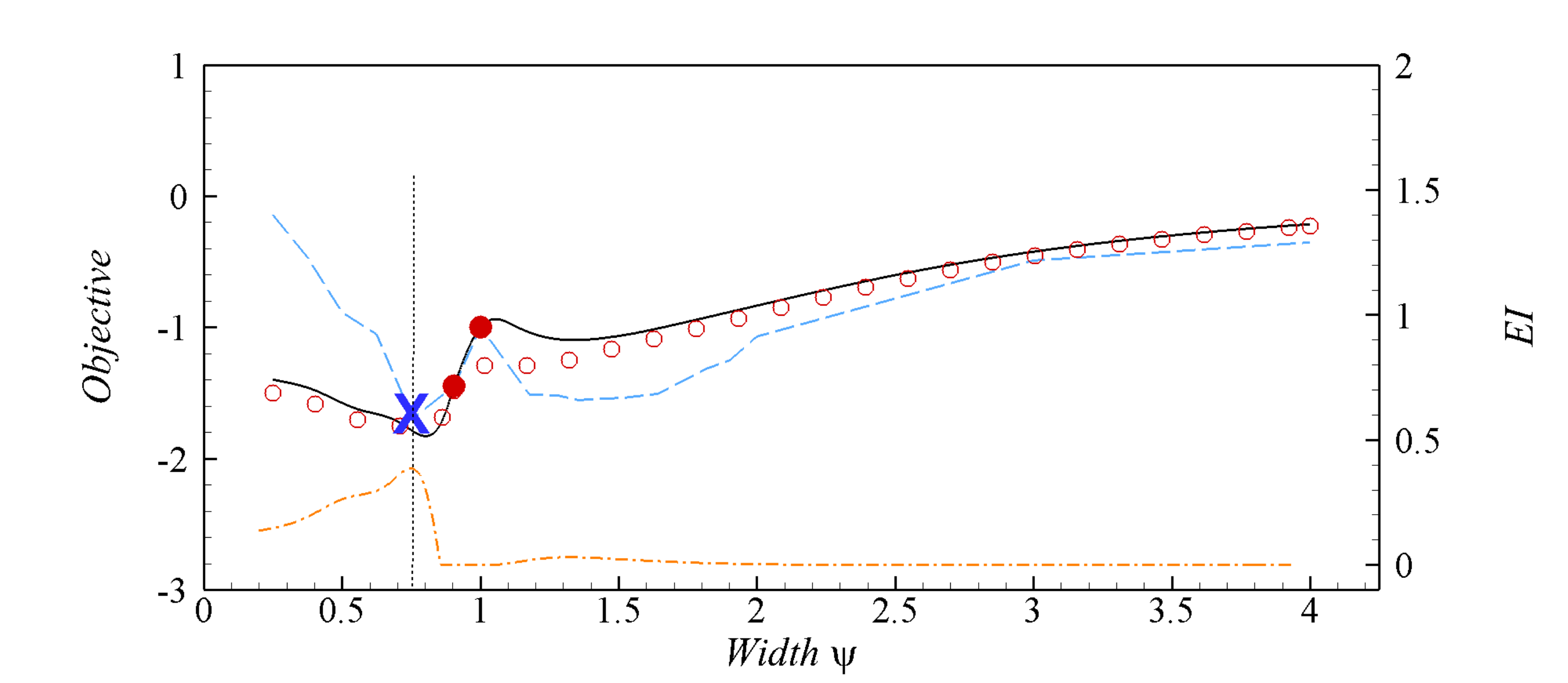}
        \caption{HK surrogate and expect improvement (EI) function at $i=1$}
	\end{subfigure}
        \\
    \begin{subfigure}[t]{0.8\textwidth}
    	\includegraphics[width=0.99\textwidth]{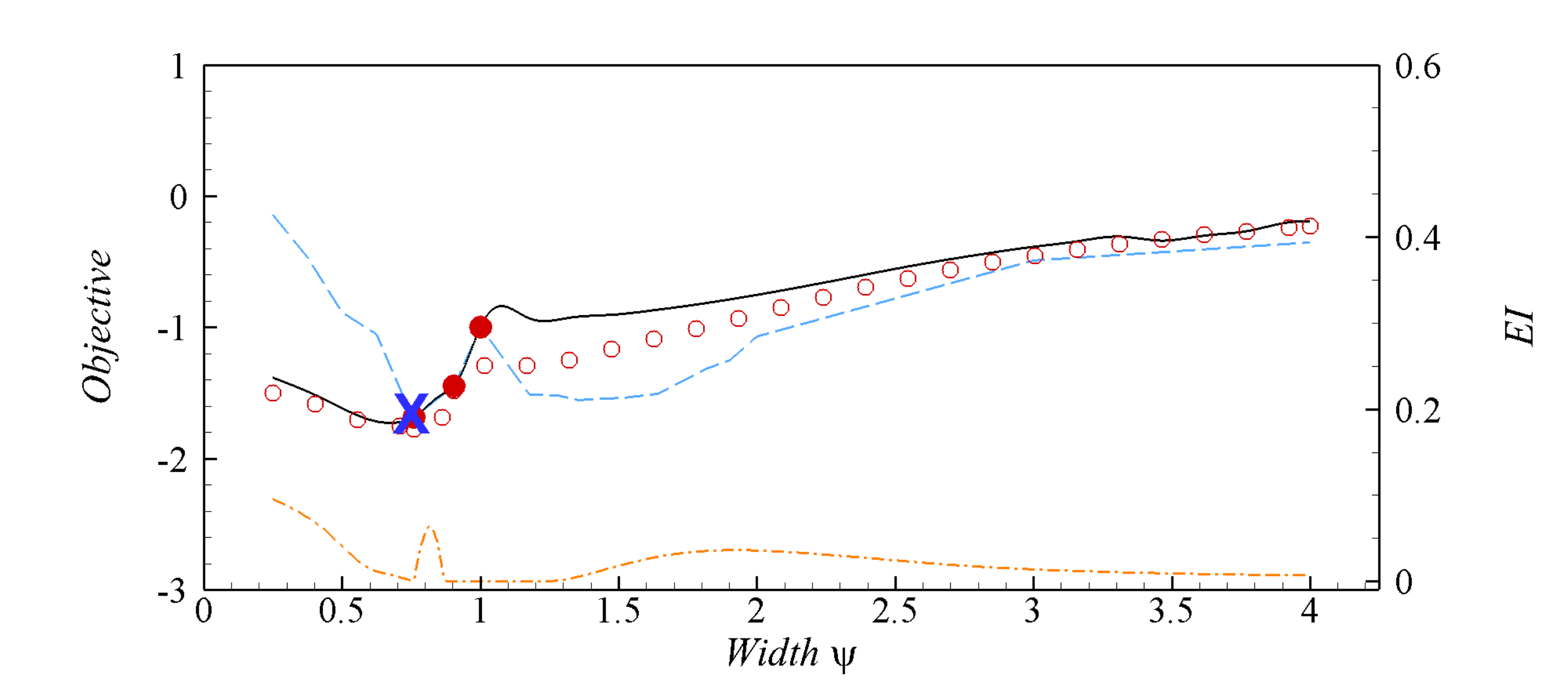}
    	\caption{HK surrogate and expect improvement (EI) function at $i=2$}
    \end{subfigure}
	\caption{Three iterations of the proposed method. (reference by LES ({\color{CornflowerBlue}- - - -}), HK ({\color{black}{\textemdash}}), EI ({\color{YellowOrange}-$\cdot$-$\cdot$-}),	hi-fidelity LES samples ({\color{red}{$\bullet$}}), and low-fidelity custom RANS ({\color{red}{$\circ$}}) )}
	\label{fig:Iter}
\end{figure}

\begin{figure}
	\centering
	\includegraphics[width=0.6\textwidth]{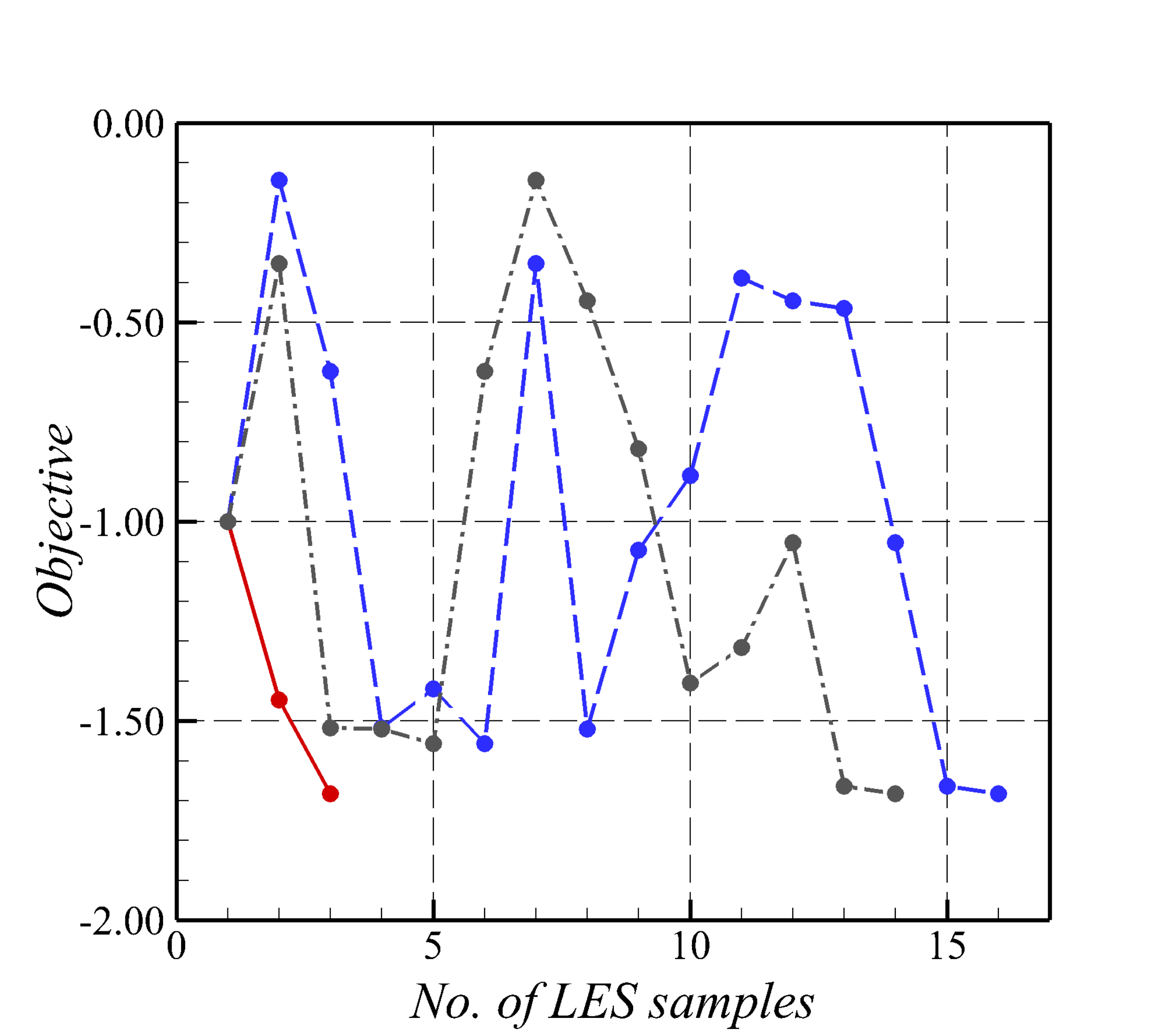}
	\caption{Convergence of bi-fidelity optimization (custom RANS as low-fidelity model({\color{red}{\textemdash}}), baseline RANS as low-fidelity model ({\color{blue}- - - -})), and single fidelity optimization using LES only ({\color{black}-$\cdot$-$\cdot$-}).}
	\label{fig:Optimum}
\end{figure}

The optimization starts, as shown in Figure~\ref{fig:Iter}(a), with one LES as the high-fidelity sample (red solid circle) and 26 custom RANS simulations as the low-fidelity samples (red hollow circle).  The initial HK response surface passes through the LES, and since there is only a single high-fidelity sample, the shape is dictated entirely by the custom RANS, and the scaling parameter $\rho$ in HK model is found to be $0.78$. The expected improvement criterion selects a second LES sample at $\psi\simeq 0.93$, Figure~\ref{fig:Iter}(a).  LES is performed at this point, and a custom closure is built, which is found by $ b^\Delta_{ij} \simeq M_{b_{ij}^{\Delta}} =  1.2 T_{ij}^{(2)}; and \hat{\textbf{R}} \simeq M_{R} = 0.45 T_{ij}^{(1)}\frac{\partial U_i}{\partial x_j}.$ Using this new closure model in RANS simulation, the low-fidelity samples near newly added sample are reevaluated. With the updated low-fidelity samples and two LES high-fidelity samples, the HK model is rebuilt, shown in Figure~\ref{fig:Iter}(b). Again the closure passes through the LES samples exactly, and the hierarchical kriging is getting more close to the true cost function in the region near the optimum. The scaling parameter $\rho$ in the HK model is found to be  $0.93$, closer to $1$ because of the better match between low- and high-fidelity objective functions.  A third iteration is shown in Figure~\ref{fig:Iter}(c).  The convergence of the procedure in terms of $J$ is shown in Figure~\ref{fig:Optimum} (red solid line).

As a comparison, we also performed a single-fidelity optimization using only the high-fidelity LES model, and a bi-fidelity optimization using the baseline RANS as the low-fidelity model. The convergence histories in terms of $J$ are shown in Figure~\ref{fig:Optimum} (black dash-dotted line and blue dashed line, respectively), where in each case we show the objective-function value of the most recently obtained sample. The convergence history of bi-fidelity optimization using the baseline RANS model as the lo-fi model, shows no improvement over the single-fidelity optimization -- and is in fact significantly more expensive due to the large number of RANS solves needed.  This is a typical example bi-fidelity methods with poor lo-fi to hi-fi correction resulting in no speed-up compared to hi-fi only. Consequently, more than $10$ LES samples are required to locate the global optimum.



Finally, the optimum solution is found at $\psi = 0.75944$ with the cost function value around $-1.68$.  In summary, the optimization has converged to within the accuracy of the high-fidelity model with only 2 LES samples, with one further sample used to verify the result.

\section{Conclusions}

In this work, a novel bi-fidelity fluid-dynamic shape optimization methodology is proposed, in which LES is used as the high-fidelity simulation tool and an automatically customized RANS turbulence closure as the low-fidelity model. The full-field LES data obtained from the high-fidelity samples are used to train a RANS model for the flow being optimized.  The SpaRTA method is used as a robust and effective way of tailoring RANS models with simple algebraic augmentation. The customized low-fidelity RANS model is successively improved as more LES data becomes available within the design space.  The two fidelities are combined in a hierarchical Kriging surrogate, and the EGO procedure is used to add new high-fidelity samples.

Given the technical complexity of this procedure, we have demonstrated it on a proof-of-concept shape optimization of a periodic-hill geometry with varying hill-width.  In this case, the method converges to the true optimum (within the tolerance of the LES simulations) in two iterations.  Shape optimization for flow-problems where the details of turbulence are important, including separated flows, junction flows, and transitional flows, may become computationally feasible with this scheme.  Future work will focus initially on application to a three-dimensional optimization problem of practical engineering interest.

\FloatBarrier
\bibliographystyle{cas-model2-names}
\bibliography{cas-refs}

\end{document}